\documentstyle[aps,preprint]{revtex}

\newcommand{\be}{\begin{equation}}
\newcommand{\ee}{\end{equation}}
\newcommand{\ba}{\begin{eqnarray}}
\newcommand{\ea}{\end{eqnarray}}
\newcommand{\ket}[1]{| {#1} \rangle}
\newcommand{\bra}[1]{\langle {#1} |}
\newcommand{\ave}[1]{\langle {#1} \rangle}
\newcommand{\mpi}{m_\pi}

\def\lsim{\mathrel{\rlap{\lower4pt\hbox{\hskip1pt$\sim$}}
    \raise1pt\hbox{$<$}}}     
\def\gsim{\mathrel{\rlap{\lower4pt\hbox{\hskip1pt$\sim$}}
    \raise1pt\hbox{$>$}}}     

\begin{document}
 
\draft

\title{The quark condensate at finite temperature}

\author{ G. Chanfray$^{(1)}$, M. Ericson$^{(1)(2)}$
and J. Wambach$^{(3)}$}
 
\address{
$^{(1)}$ IPN-Lyon, 43 Bd. du 11 Novembre 1918,
F-69622 Villeurbanne C\'{e}dex, France.\\
$^{(2)}$ CERN, Geneva, Switzerland.\\
$^{(3)}$ Institut f\"{u}r Kernphysik, Technische Hochschule
Darmstadt,\\ Schlo{\ss}gartenstra{\ss}e 9, D-64289 Darmstadt, Germany.}

\date{July 15, 1996}
\maketitle

\begin{abstract}
The temperature evolution of the quark condensate is studied using three 
different methods. In the spirit of a many-body approach we make an 
expansion in the scalar density up to second order. Our result is 
consistent with chiral perturbation theory to two-loop order.

\end{abstract}


\section{Introduction}
Chiral symmetry, which is an approximate symmetry of QCD
in the light-flavor sector, is spontaneously
broken in the QCD vacuum. A topic of interest for nuclear physics is  
the restoration of chiral symmetry, at high temperature and/or large
baryonic density. The approach to restoration shows up as a decrease 
of the magnitude of the quark condensate, which is the order parameter of 
the spontaneous breaking. Within a hadronic picture the physical origin 
of this restoration is very simple in nature. Each hadron contributes a 
certain amount to the decrease, governed by its sigma commutator
$\Sigma_h$. For an assembly of independent hadrons, their effects add. This 
leads to the following expression for the evolution with temperature of the 
condensate
 
\begin{equation}
\frac{\ave{q\bar q}_T}{\ave{q \bar q}_0}=
1-\sum_h \frac {\Sigma_h \rho^s_h(T)}
{f^2_\pi\, m^2_\pi}
\label{IND}
\end{equation}
where the sum extends over all hadron species present in the medium and
$\rho^s_h$ is the corresponding scalar density. In a baryon free regime
(zero baryonic chemical potential) the scalar density is produced by 
thermally excited hadrons. At low temperature eq.~(\ref{IND}) 
is the leading contribution. Deviations arise from the interaction between 
the hadrons and the major problem is to evaluate their effect in a way 
which is consistent with chiral symmetry. In this work we shall investigate
the temperature evolution of several quantities linked to chiral symmetry, 
namely the quark condensate, the pion mass and the weak pion decay constant.
The results that we obtain in the chiral Lagrangian approach have already
been given in chiral perturbation theory \cite{GaLe,GeLe}. Our approach 
provides alternative methods suitable for incorporating other hadron species,
especially vector mesons, in a consistent way.
 
For temperatures less than the pion mass the presence of mesons other than 
pions can be ignored, since their thermal abundance is small. It is also 
sufficient to expand to second order in the density. 
 
To respect chiral symmetry we start from the standard nonlinear chiral 
Lagrangian, including the explicit chiral symmetry breaking piece, namely
 
\begin{equation}
{\cal L}= {f^2_\pi \over 4}\,{\rm Tr}\partial_\mu U \partial^\mu U^\dagger\,
+\,{1\over 4} f^2_\pi m^2_\pi\, {\rm Tr}(U+U^\dagger).
\end{equation}
The $SU(2)$-matrix $U$ is usually expressed in terms of the canonical pion 
field $\vec \Phi$ as
\begin{equation}
U=\exp\left(i\vec \tau.\hat \Phi F(X)\right)
\end{equation}
where $F(X)$ is an arbitrary odd polynomial of
$X=(\Phi^2/ f^2_\pi)^{1/2}$
 
\begin{equation}
F(X)=X+\alpha X^3+\beta X^5+.....
\label{EXPA}
\end{equation}

\noindent
Different choices of the function $F(X)$ have been introduced in
the literature, corresponding to different values of the parameters $\alpha$ and
$\beta$. For instance, the simple choice $F(X)=X$ is commonly used in chiral
perturbation expansions. The choice of Weinberg is $ F(X)=
\arcsin [X/(1+X^2/4)]$ \cite{Wein}.
On the other hand $F(X)= \arcsin X$ gives the PCAC field, satisfying 
$\partial_\mu A^\mu_i=-m_\pi f^2_\pi \Phi_i$. Physical observables
should not depend on the particular function $F(X)$. In practice, however,
a particular choice is made and some approximations are needed. It is thus
important to check the independence of the results on $F(X)$. 

In the case of a cold baryonic medium the independence on the representation
has been established recently by Delorme et al. \cite{Delo} and by 
Kirchbach and Wirzba \cite{KiWi}. It is one of the purposes of the present
paper to explicitly check this property for the quark condensate in a hot 
baryon-free environment. A baryon-free zone may possibly be realised 
experimentally at RHIC and the LHC and is certainly present in the early 
universe. To derive the temperature dependence of the quark condensate we 
will use three methods.
The first (sect.1) relies on the grand potential method introduced in
ref.~\cite{BuWa}. In the second method (sect.2) we evaluate the 
thermal expectation value of the chiral symmetry breaking piece of the 
Lagrangian. Both methods demonstrate the expected independence of the result
on the particular choice of $F(X)$. Once this independence is proven we 
will introduce a third method (sect.~3) which requires the validity of the 
PCAC relation, {\it i.e.} corresponds to a particular choice of $F(X)$. In 
this case the condensate can be derived as the scattering amplitude of soft
pions in the system. The condensate is then expressed in terms of the 
self energy for soft pions. In this derivation we will prove the validity 
of the Gell-Mann-Oakes-Renner relation (GOR) in a hot medium, extending the 
previous derivation of Chanfray et al. \cite{Chan} which was restricted to 
zero temperature.
 
\section{The grand potential method}

To obtain an expression for the chiral condensate ratio,
$\ave{\bar qq}_T/\ave{\bar qq}_0$, in terms of pionic degrees of freedom
we can use the fact that, according to the Feynman-Hellmann theorem,
$\ave{\bar qq}_T$ is related to the derivative of the free-energy density
(the grand potential) $\tilde\Omega(T)$ with respect to the bare 
quark mass $m$ as

\be
\ave{\bar q q}_T=\partial\tilde\Omega(T)/\partial m
\ee
(we ignore the fact that up and down quark masses not the same).
Denoting the difference in free energy density as 
$\Omega(T)=\tilde\Omega(T)-\tilde\Omega(0)$ and using the Gell-Mann Oakes 
Renner relation (GOR) in the vacuum \cite{GeOR}

\be
m_\pi^2 f_\pi^2=\,-2m\ave{\bar qq}_0\,=\,-m{\partial\tilde\Omega(0)\over
\partial m}
\label{GOR}
\ee
one immediately derives that

\be
{\ave{\bar qq}_T\over\ave{\bar qq}_0}=
1-{1\over f_\pi^2}{\partial\Omega(T)\over\partial\mpi^2}
\label{CCOND}
\ee
This expression is identical to that of ref.~\cite{GeLe} if one
ignores small corrections to $f_\pi$ and $\mpi$ of order $m$.
\smallskip
For a free pion gas the grand potential is
\be
\Omega(T)={3 T\over 2 \pi^2}\,\int_0^\infty\,dk\,k^2\,\rm{ln}
\left(1-e^{-\omega_k/T}\right);\qquad \omega_k=\sqrt{m_\pi^2+{\bf k}^2}.
\ee
Taking the derivative with respect to the pion mass and using (\ref{CCOND})
leads to

\be
{\ave{\bar qq}_T\over\ave{\bar qq}_0}=
1-{m_\pi/2\over f_\pi^2m_\pi^2}\rho^s_\pi(T)
\label{CCONF}
\ee
with  

\begin{equation}
\rho^s_\pi(T)=3\,\int {d{\bf k}\over (2\pi)^3}\, {m_\pi\over  \omega_k}\, 
{1\over \exp (\omega_k/T)-1}.
\end{equation}
Keeping in mind that the pion sigma commutator is $\Sigma_\pi=m_\pi/2$ (with
 nonrelativistic normalization) and that $\rho^s_\pi$ is the scalar density
of pions, we recover the independent-particle expression of eq.~(\ref{IND}) 
for a gas of free pions.

In the evaluation of the free energy, $\Omega(T)$, for interacting pions we 
start from the well-known expression for interacting boson systems at finite 
temperature \cite{LuWa,CaPe}. In terms of the pion Green's function
 
\be
D(\omega,{\bf k} ,T)=(\omega^2-m_\pi^2-{\bf k} ^2
-\Pi(\omega,{\bf k} ,T))^{-1},
\ee
the grand potential is given by

\ba
\Omega(T)=\Omega'(T)-{3\over 2}\int
{d{\bf k}\over (2\pi)^3}\int {d\omega\over \pi }\,f(\omega)\,
{\rm Im} \biggl \{
\rm{ln}[-D^{-1}(\omega,{\bf k} ,T)]
+D(\omega,{\bf k} ,T)\Pi(\omega,{\bf k} ,T)\biggr \}
\label{OMEG}
\ea
where $f(\omega)=\left(\exp(\omega/T)-1\right)^{-1}$ is the Bose factor and 
$\Omega'(T)$ is the sum of all contributions from 'skeleton diagrams' 
representing the perturbation expansion of $\Omega$. These are evaluated by 
using full single-particle Green's functions rather than bare ones.
Furthermore

\be
\delta \Omega /\delta \Pi =0
\label{STAT1}
\ee
and
\be
\delta \Omega'/\delta D =\Pi
\label{STAT2}
\ee
(see refs.\cite{LuWa,CaPe,AbGD}). 
Since the physical pion mass enters the propagator, $D$, the self energy
$\Pi (\omega ,{\bf k} ,T)$ is the difference

\be
\Pi(\omega,{\bf k},T)=\tilde\Pi(\omega,{\bf k},T)-\tilde\Pi(\omega,{\bf k},0),
\ee
whereby the infinite contributions to the mass operator
 $\tilde\Pi(\omega,{\bf k},0)$
due to vacuum fluctuations are removed.
Consequently also $\Omega(T)$ is finite and no regularization of the 
loop integrals is needed.

The key quantity in eq.~(\ref{OMEG}) is the pion self energy $\Pi$. To 
calculate it we use the nonlinear $\sigma$-Model Lagrangian defined
in eq.~(2). To lowest order, the interaction part takes the form

\begin{equation}
{\cal L}_{int}= {1\over f^2_\pi} \bigg[ -m^2_\pi \left(\alpha-{1\over 24}
\right)
(\vec \Phi. \vec \Phi)^2\,+\,\left(\alpha-{1\over 6}\right) \vec \Phi.\vec\Phi\,
\partial^\mu \vec \Phi. \partial_\mu \vec\Phi\,+\,
\left(2 \alpha +{1\over 6}\right) \vec \Phi. \partial^\mu \vec \Phi\,
 \vec \Phi. \partial_\mu \vec \Phi \bigg] .
\label{LINT}
\end{equation}
The pion self energy $\Pi$ contains single-pion and three-pion contributions
(see Fig.~1a and 1b)

\be
\tilde\Pi(\omega,{\bf k} ,T)=
\tilde\Pi(\omega,{\bf k} ,T)^{1\pi}+\tilde\Pi(\omega,{\bf k} ,T)^{3\pi} .
\ee
In the present approach and for the temperature range of interest 
($T\le \mpi$), only the single-pion contribution is needed. This is 
equivalent to the Hartree approximation and leads to~: 

\begin{equation}
\Pi(\omega,{\bf k},T)=\lambda\ j(T) \bigg[ \left(80\alpha-{10\over 3}\right)
m^2_\pi
\,-\,\left(40\alpha-{8\over 3}\right) (\tilde m^2_\pi+
\omega^2-{\bf k}^2)\bigg]
\label{SELF1}
\end{equation}

\noindent
where $\lambda\equiv 1/6f^2_\pi$ and $j(T)$ is the thermal loop integral 
given by

\begin{equation}
j(T)=3\,\int {d{\bf k}\over (2\pi)^3}\, {1\over 2 \tilde\omega_k}\, {1\over
\exp (\tilde\omega_k/T)-1}.
\end{equation}
Here $\tilde \omega_k=(\tilde m_\pi^2 +{\bf k}^2)^{1/2}$ is the in-medium 
pion dispersion relation with an effective mass $\tilde m_\pi$. Its value 
is obtained by solving the following equation~:

\begin{equation}
\tilde m^2_\pi=m^2_\pi +\Pi(\tilde m_\pi,\, {\bf k}=0,T)
\end{equation}
which yields
 
\be
\tilde\mpi^2={1+(80\alpha-10/3)\lambda j\over 
1+(80\alpha-16/3)\lambda j}\,\mpi ^2
\label{EFFM}
\ee
Since $\tilde\mpi^2$ is proportional to the bare mass it vanishes in the
chiral limit $\mpi\to 0$, at all temperatures as required by the
Goldstone theorem. Notice that, for 'on-shell' pions, the 
$\alpha$-dependence of the self energy disappears, 
as it should, and $\Pi_{os}=2\lambda j$. The pion propagator 
$D(\omega,{\bf k},T)$ can now be written as

\be
D(\omega ,{\bf k} ,T)
=\gamma (\omega ^2 -{\bf k}^2 -\tilde \mpi ^2 )^{-1},
\label{PROP2}
\ee
In addition to the effective mass it also contains a temperature dependent 
residue \cite{BuWa}
 
\begin{equation}
\gamma(T)=\biggl [ 1+\left(40 \alpha-{8\over 3}\right) \lambda j\biggr ]^{-1}
\label{RESI}
\end{equation}
which arises from the energy dependence of the $\pi\pi$-interaction given
in (\ref{LINT}).

Finally we need to evaluate the interaction part $\Omega '$ which takes 
the form
\ba
\Omega '={\lambda\over 2}\int {d{\bf k} \over (2\pi )^3}
\int {d{\bf p} \over (2\pi )^3}
\int _{-\infty} ^{\infty} {d\omega\over \pi} f(\omega)
Im D(\omega ,{\bf k} ,T)
\int_{-\infty}^{\infty}{d\eta \over \pi} f(\eta)
 Im D(\eta ,{\bf p} ,T)\nonumber\\
\times\left(80\alpha-{10\over 3}\right)m^2_\pi
\,-\,\left(40\alpha-{8\over 3}\right)\biggl [ (\omega^2-{\bf k}^2))
+(\eta^2-{\bf p}^2))\biggr ]\, .
\ea
Once $\Omega(T)$ is determined one can find the chiral condensate ratio  
from eq.~(\ref{CCOND}). When taking the derivative one should note that 
$\Omega$ depends on the free pion mass $\mpi$, both explicitly and through
the self energy $\Pi$. Due to the stationarity conditions (\ref{STAT1}) and 
(\ref{STAT2}) the $\mpi$-dependence of $\Pi$ does not influence 
$\partial \Omega /\partial \mpi ^2$ and we obtain

\be
{\partial \Omega\over \partial \mpi ^2}= d(T)-{\lambda
\over 2}(80\alpha-10/3)d^2(T).
\ee
where $d(T)\equiv\gamma(T) j(T)$ which is identical to the result of
ref.~\cite{BuWa} for $\alpha=-1/12$ (the 'Weinberg' choice for F(X) 
\cite{Wein}). Hence, with eq.~(\ref{CCOND}),

\be
{\ave{\bar qq}_T\over\ave{\bar qq}_0}=
1-6\lambda d(T)-(240\alpha-10)(\lambda d(T))^2
\ee
At first sight, this expression has an unwanted $\alpha$-dependence. 
This model dependence disappears, however, if we expand the residue 
given in eq.~(\ref{RESI}) to first order in the coupling constant $\lambda$.
This gives the evolution of the condensate 
to second order in $\lambda$, independent of $\alpha$
 
\be
{\ave{\bar qq}_T\over\ave{\bar qq}_0}=
1-6\lambda j(T)-6(\lambda j(T))^2
\label{EXP2}
\ee 
where the first-order term $6 \lambda j$ is just the contribution from a 
free pion gas (see eq.~(\ref{CCONF})), which dominates.
The expression (\ref{EXP2}) represents, in the spirit of a many-body approach,
an expansion in the density. We can check its consistency with chiral
perturbation theory.  In the chiral limit ($m_\pi \to 0$) where  $j$
can be calculated analytically ($j=T^2/8$), we recover the chiral
perturbation result to order $T^4$ 
 
\begin{equation}
{\ave{\bar qq}_T\over\ave{\bar qq}_0}= 1\,-\,{T^2\over 8 f^2_\pi}\,-\,
{T^4\over 384 f^4_\pi}.
\end{equation}
of Gasser and Leutwyler \cite{GaLe}.

\section{Expectation value of the chiral symmetry breaking Hamiltonian}

The condensate ratio can also be obtained directly from the in-medium 
expectation value of the symmetry breaking Hamiltonian of the
nonlinear sigma model (eq.~(2))
 
\begin{equation}
H_{\chi SB}=-{1\over 4} m^2_\pi f^2_\pi tr(U+U^\dagger)= -m^2_\pi f^2_\pi 
\cos F(X). 
\end{equation}
Using the GOR in vacuum (eq.~(\ref{GOR})) we obtain

\begin{equation}
{\ave{\bar qq}_T\over\ave{\bar qq}_0}=\ave{\cos \left(F(X)\right)}_T
\label{cosF}
\end{equation}
where $\ave{..}_T$ on the {\sl rhs} denotes a thermal expectation value.
In order to be consistent with the second-order expansion in 
expression (\ref{EXP2}) we have to evaluate the thermal 
expectation value to fourth order in the pion field~:

\begin{equation}
{\ave{\bar qq}_T\over\ave{\bar qq}_0}=1-{1\over 2}
\ave{{\Phi^2\over f^2_\pi}}_T-
\left(\alpha-{1\over 24}\right)
\ave{\Phi^2 \Phi^2 \over f^4_{\pi}}_T
\label{ratio}
\end{equation}
In order to evaluate the expectation values explicitly,
the pion field is expanded in terms of creation and annihilation operators
$B^\dagger_{i,{\bf k}}$ and $B_{i,{\bf k}}$ for the quasi pions, specified
by the propagator given in (\ref{PROP2}) 
 
\begin{equation}
\Phi_i({\bf r},t)=\gamma^{1\over 2}\,
\int {d{\bf k}\over \left( (2\pi)^3\, 2 \omega_k\right)^{1/2}}\,
\left(B_{i,{\bf k}}(t) +B^\dagger_{i,-{\bf k}}(t)\right)\, 
e^{i{\bf k}.{\bf r}}
\label{FIELD}
\end{equation}

\noindent
The presence of the $\gamma^{1/2}$ factor ensures
that the form (\ref{PROP2}) of the pion propagator $D$ is recovered once 
the time dependence $B_{\bf k}(t)=B_{\bf k}\, \exp(-i\tilde \omega_k t)$ 
is introduced. We have verified that the field operator (\ref{FIELD}) obeys 
canonical commutation relations. The expectation value of $\ave{\Phi^2}_T$
is calculated by normal ordering with respect to physical vacuum and yields
 
\begin{equation}
\ave{\Phi^2}_T=\gamma\, \int {d{\bf k}\over (2\pi)^3\, 2\tilde \omega_k}\,
\sum_{i=1}^3 \,
\ave{B_{i{\bf k}}^\dagger \, B_{i{\bf k}}}_T= 2\gamma\, j
\label{PHI2}
\end{equation}
Similarly $\ave{\Phi^2 \Phi^2}_T$ is obtained by using Wick's theorem and
gives
 
\begin{equation}
\ave{\Phi^2 \Phi^2}_T={5\over 3}\, \ave{\Phi^2}_T^2=\, {20 \over  3} \gamma^2
j^2
\label{PHI4}
\end{equation}
Expanding the residue $\gamma$, defined in (\ref{RESI}), for both 
(\ref{PHI2})
and (\ref{PHI4}) to first order in $\lambda j$  and inserting in 
(\ref{ratio}), we recover the expression (\ref{EXP2}) of the previous 
section. In the next section we will introduce a third method which is based
on the use of the PCAC pion field.
 
\section{Soft-pion method}
In this section, the values of the parameters $\alpha$ and $\beta$ in the
expansion of $F(X)$ (eq.~(\ref{EXPA})) are chosen such that the PCAC 
relation

\be 
\partial_\mu A^\mu_i(x)= -f_\pi\, m_\pi^2\,\Phi_i(x)
\ee
is fulfilled. This corresponds to the values $\alpha=1/6$ and $\beta= 3/40$.
In the previous sections, however, it has been shown that the condensate 
ratio, to second order in the density, is independent on these parameters, 
which legitimates this third approach. The aim here is to obtain the 
condensate ratio from its relation to the self energy for soft pions.
In deriving such a relation we will show the validity of the in-medium 
version of the GOR which relates the effective pion mass, the effective pion 
decay constant and the in-medium quark condensate. This has been previously 
shown to hold in a dense but cold hadronic medium by Chanfray {\sl et al.} 
\cite{Chan}.

We start from the QCD operator identity
 
\begin{equation}
\bigg[\, Q^5_i(0)\, , \, \partial_\mu A^\mu_i(0)\,\bigg]=-i\,2 m \,
\bar q q(0).
\end{equation}

\noindent
Taking the thermal expectation value of this relation and using
closure on the {\sl lhs} one has
 
\begin{equation}
\int d{\bf r}\,\sum_{n,m}\,{e^{-\beta E_n} \over Z}\,
\left(\bra{n} A_i^0({\bf r},0)\ket{m}\bra{m}\partial_\mu A^\mu_i(0)\ket{n}
-\bra{n}\partial_\mu A^\mu_i(0)\ket{m}\bra{m} A_i^0({\bf r},0)\ket{n}\right)
\nonumber\\
\end{equation}
\begin{equation}
=-2i m  \ave{\bar q q}_T
\label{com}
\end{equation}
where $Z$ is the partition function.
After spatial integration, translational invariance requires that 
${\bf p}_n= {\bf p}_m$.

The above equation is equivalent to the following energy-weighted sum rule
\be
2m \ave{\bar q q}_T\,=-\,\sum_{n,m}\, {e^{-\beta E_n}\over Z}\,
2(E_m-E_n)\,|\bra{m} A^0_i(0)\ket{n}|^2, 
\label{SUMR}
\ee
It is now possible to derive a generalized GOR relation at finite 
temperature. For this purpose we assume that the sum rule (\ref{SUMR}) is 
saturated by exciting a single quasi pion, $\tilde\pi$, 
(of mass $\tilde m_\pi$) from the heat bath, {\sl i.e.}
\be
2m \ave{\bar q q}_T\,=-\,\sum_n\,{e^{-\beta E_n}\over Z}\, 2\,\tilde
m_\pi\,|<n+\tilde \pi|\,A^0_i(0)\,|n>|^2
\ee
It is natural to introduce a thermally averaged pion decay constant, $\tilde
f_\pi$, through
\be
\sum_n\,{e^{-\beta E_n}\over Z}\,|<n+\tilde \pi|\,A^0_i(0)\,|n>|^2=\, 
{(\tilde f_\pi\, \tilde m_\pi)^2 \over 2 \tilde m_\pi}
\ee
The generalized GOR relation follows immediately
\be
2m \ave{\bar q q}_T\,=\,-\tilde f^2_\pi\, \tilde m^2_\pi.
\label{GGOR}
\ee
Notice that Lorentz invariance is broken by the very concept of
temperature. Therefore the renormalization of $f_\pi$ from the GOR relation
(\ref{GGOR}) concerns only the time component of the axial current.

Coming back to the energy-weighted sum rule (\ref{SUMR}), we 
introduce the PCAC interpolating pion field through
$\Phi_i(x)=-\partial_\mu A^\mu_i(x)/ f_\pi m^2_\pi$.      
Taking the matrix element of this PCAC identity between states $\ket{m}$ and
$\ket{n}$ yields
 
\begin{equation}
i(E_n-E_m)\,\bra{n}A^0_i(0)\ket{m}=- f_\pi m^2_\pi \,\bra{n}\Phi_i(0)\ket{m}
\label{PCAC0}
\end{equation}
and leads to an alternative form of the sum rule
 
\begin{equation}
2 m\,\ave{\bar q q}_T=-2 f^2_\pi m^4_\pi\,\sum_{n}
{e^{-\beta E_n}\over Z}
\,\sum_m
{|\bra{m} \Phi_i(0)\ket{n}|^2 \over E_m-E_n}.
\label{sumrul}
\end{equation}
This relation links the condensate to the axial polarizability of the 
thermal bath. On the other hand it is a simple matter to show that the 
thermal pion propagator $D(0,0,T)$, calculated at the soft pion point 
$\omega=0,\, {\bf k}=0$, is
 
\begin{eqnarray}
D(0,0,T)&\,=\,& -\big(m^2_\pi\,+\,\Pi(0,0,T)\big)^{-1}
\nonumber\\
    &\,=\,& \int dt\, \int d{\bf r}\,\ave{(-i)\left[ \Phi_i({\bf r},t)\,
,\,\Phi_i(0)\right]}_T\,
    \theta(t)
\nonumber\\
    &\,=\,& -2\,\sum_{n,m}{e^{-\beta E_n}\over Z}\,\sum_m
{|\bra{m} \Phi_i(0)\ket{n}|^2 \over E_m-E_n}
\label{prop}
\end{eqnarray}
Comparing eqs.~(\ref{sumrul}) and ({\ref{prop}), we obtain 
 
\begin{equation}
{\ave{\bar qq}_T\over\ave{\bar qq}_0}=
\biggl (1\,+\,{\Pi(0,0,T)\over m^2_\pi}\biggr )^{-1}
\label{ratio2}
\end{equation}
This is the generalization of a result already obtained for nuclear matter
by M. Ericson \cite{Eric}, which incorporates the coherent rescattering of 
the soft pion in an optical potential coinciding with the soft-pion 
self energy $\Pi(0,0,T)$. 

In order to evaluate the condensate, we need the expression of the self 
energy to two-loop order, including in particular the contribution from the 
diagram of Fig.~1c. These were not necessary in our previous approach 
(\ref{SELF1}). The two-loop contribution requires to expand 
the Lagrangian to sixth order in the pion field.  Both parameters $\alpha$  
and $\beta$ enter in this case. The result for the self-energy is
 
\begin{eqnarray}
\Pi(\omega ,\,{\bf k},T)&\,=\,&\bigg({5\over 6}\ave{X^2}_T+
{35\over 24}\ave{X^2}_T^2\bigg) \,m^2_\pi
-\bigg({1\over 3}\ave{X^2}_T +{10\over 9}\ave{X^2}^2_T\bigg) \tilde m^2_\pi
\nonumber\\
&\,-\,&\bigg({1\over 3}\ave{X^2}_T +{5\over 9}\ave{X^2}^2_T\bigg) 
\,(\omega^2-{\bf k}^2)
\label{self}
\end{eqnarray}
where $\ave{X^2}_T=12 \lambda j$ represents the thermal loop integral of 
quasi pions. In expression (\ref{self}) we take for consistency the 
effective mass to first order, {\sl i.e.} 
$\tilde m^2_\pi=m^2_\pi(1+2\lambda j)$.  We obtain the self energy for soft
pions from (\ref{self}) and when inserting its value in (\ref{ratio2}) we 
recover, to second-order in $\lambda j$, our previous result of 
eq.~(\ref{ratio}).
Notice that the expression (\ref{self}) for $\Pi(\omega,{\bf k},T)$
is valid at the level of an extended Hartree approximation. To this order 
there is an additional contribution corresponding to a pion fluctuating in 
three pions (see Fig.~1b) which, however vanishes at the soft pion point. 
Therefore it does not affect the soft-pion self energy and hence the 
condensate. However it enters in the residue $\gamma$ to second order in
the coupling constant $\lambda$. Since we did not calculate this term, our
residue and hence our expression (21) for the pion mass are only valid to 
first order. This is also the case for the pion decay constant 
$\tilde f_\pi$. From  the GOR relation we obtain 
$\tilde f_{\pi}=f_{\pi}(1-4\lambda j)$, in agreement with ref.~\cite{GaLe}.

In summary we have studied the evolution with temperature of the quark
condensate in the framework of a many-body approach, considering thermal
excitations of pions. The $\pi\pi$ interaction is taken from the nonlinear
sigma model. Our expansion parameter is the scalar density of the quasi 
pions and we work up to second order. We ave shown in two different 
approaches that, up to this order, the condensate is independent of the
transformation of the pion field as expected. In line with ref. \cite{Eric}
we have used a third method, based on PCAC, to derive the condensate which 
is related  to the self energy of soft pions. In the chiral limit our results 
are consistent with chiral perturbation theory, to two-loop order.
The next stage will be to 
introduce the rho meson exchange in the $\pi\pi$ interaction in a way
which preserves the chiral properties of this interaction.

\newpage

\noindent
{\bf Acknowledgement}:
We thank J. Delorme for fruitful discussions.
M. Ericson acknowledges support from the Humboldt foundation.
This work is also supported in part by the National
Science Foundation under Grant No. NSF PHY94-21309.

\newpage
\begin{center}
{\Large \sl \bf Figure Captions}
\end{center}
\vspace{1.0cm}

\begin{itemize}
\item[{\bf Fig.~1}:] The pion self energy including the tadpole (a) 3-$\pi$
intermediate states (b) and the two-$\pi$ loop contribution (c).

\end{itemize}

\newpage

\begin{figure}
\vspace{22cm}
\includegraphics{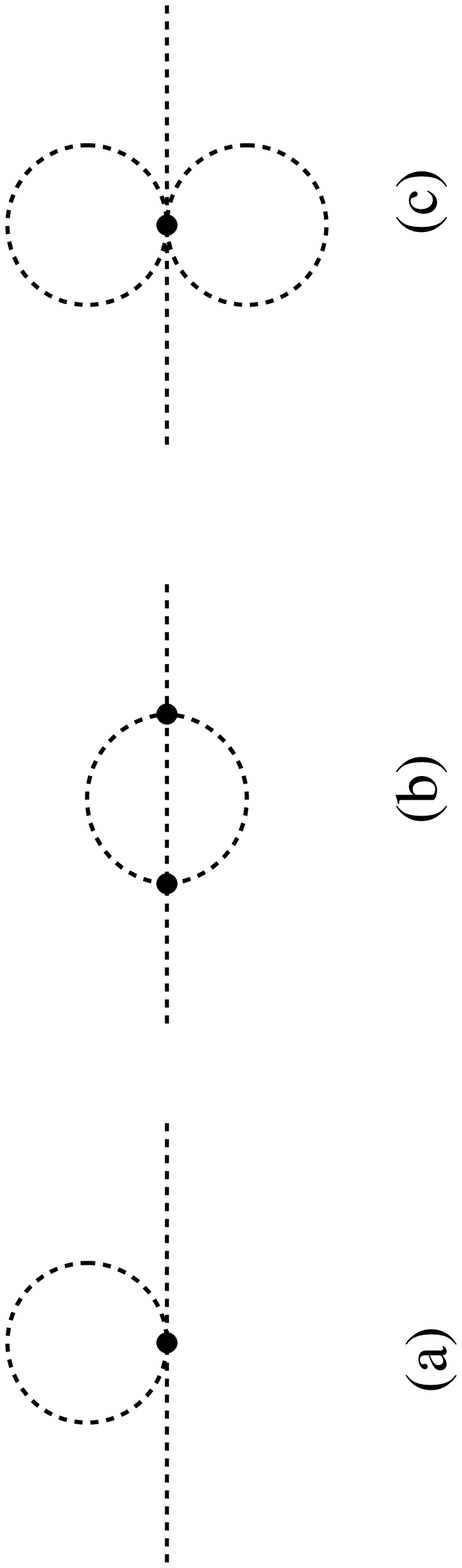}

\end{figure}

\end{document}